# Stability and Adsorption Properties of Electrostatic Complexes : Design of Hybrid Nanostructures for Coating Applications

Ling Qi, Jean-Paul Chapel, Jean-Christophe Castaing, Jérôme Fresnais and Jean-François Berret

*Complex Fluid Laboratory, UMR CNRS/Rhodia 166, Rhodia North-America, R&D Headquarters CRTB, 350 George Patterson Blvd., Bristol, PA 19007 USA, France and Matière et Systèmes Complexes, UMR 7057 CNRS Université Denis Diderot Paris-VII, Bâtiment Condorcet, 10 rue Alice Domon et Léonie Duquet, 75205 Paris, France*

RECEIVED DATE (automatically inserted by publisher); Author E-mail Address: jean-francois.berret@univ-paris-diderot.fr

We report the presence of a correlation between the bulk and interfacial properties of electrostatic coacervate complexes. Complexes were obtained by co-assembly between cationic-neutral diblocks and oppositely charged surfactant micelles or 7 nm cerium oxide nanoparticles. Light scattering and reflectometry measurements revealed that the hybrid nanoparticle aggregates were more stable both through dilution and rinsing (from either a polystyrene or a silica surfaces) than their surfactant counterparts. These findings were attributed to a marked difference in critical association concentration between the two systems and to the frozen state of the hybrid structures.

The design of functional molecular architectures and materials has attracted much attention during the last decade. In particular, the controlled association of polymers and nanoparticles using covalent [1-3] and non-covalent [4-6] binding has appeared as a promising way to combine organic and inorganic moieties at a nanoscospic level. The development of hybrid nanostructures was also stimulated by industrial and biomedical applications, especially by applications in the realm of coating technologies. In order to modify the surface properties of materials, inorganic nanoparticles with unique physical features (such as magnetic, fluorescent, UV-absorbent, high dielectric constant or catalytic properties) are actually crucial ingredients. In association with macromolecules, these nanoparticles could be also used for coatings with improved stability and performances (wetting, anti-biofouling).

In 2004, Cohen Stuart and coworkers have shown that electrostatic complexes made from oppositely charged polyelectrolytes could be effectively adsorbed on hydrophilic and hydrophobic surfaces [7-9]. Experiments were conducted at the liquid-solid interfaces on electrostatic core-shell complexes resulting from the self-assembly of an anionic-neutral copolymer and a short cationic homopolymer. On silica and polystyrene substrates, it was found that the cores of the aggregates adsorbed on the surface whereas the shell formed a brush on the top of it. These authors also confirmed the stability of the deposited layer upon rinsing and its repellent effect with respect to proteins [7]. The approach followed in the present communication aimed to extend these measurements to two new types of electrostatic systems, namely to organic and hybrids coacervate complexes. Here, we show the existence of a correlation between the bulk and adsorption properties of surfactant/copolymer (organic) and nanoparticle/copolymer (hybrid) complexes. Using Stagnation Point Adsorption Reflectometry (SPAR), organic and hybrid complexes were found to adsorb readily on hydrophilic and hydrophobic substrates. However, upon rinsing the organic complexes were shown to disassemble and finally desorb from the solid surface, whereas the hybrids remained. These findings were interpreted in terms of a critical association concentration which is much higher for organic systems than for the hybrids.

For electrostatic complexation, we have used polyelectrolyte-neutral block copolymers abbreviated in the following as $PTEA_{11K}$-$b$-$PAM_{30K}$, where PTEA stands for poly(trimethylammonium ethylacrylate methylsulfate), PAM for poly(acrylamide) and the indices represent the molecular weight of each block. In aqueous solutions at neutral pH, the chains are dispersed and in a state of unimers. Light scattering performed in the dilute regime have revealed a molecular weight $M_W$ = 35 000 g mol$^{-1}$ and an hydrodynamic diameter $D_H$ = 11 nm [10,11].

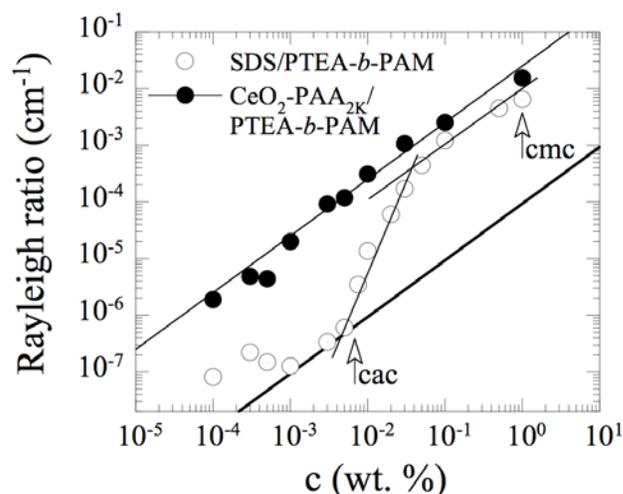

*Figure 1* : Concentration dependences of the Rayleigh ratios obtained by light scattering for SDS/$PTEA_{11K}$-$b$-$PAM_{30K}$ and for $CeO_2$-$PAA_{2K}$/$PTEA_{11K}$-$b$-$PAM_{30K}$ mixed systems. The critical association concentration for SDS/$PTEA_{11K}$-$b$-$PAM_{30K}$ ($c_{cac}$ = 6.8×10-3 wt. %) and the critical micellar concentration ($c_{cmc}$ = 1.1 wt.%) for SDS are indicated by arrows. Note that the cmc was estimated from the actual surfactant content present in the solution (1.1 wt. % in total concentration represents here 0.24 wt. % or 8.3 mM in SDS [19]). The thick straight line depicts the intensities calculated assuming that surfactants and polymers are not associated.

Complexation schemes were realized using cerium oxide nanoparticles or surfactant micelles that were of opposite charge with respect to that of the polymers. Anionic sodium dodecylsulfate (SDS) were purchased from Sigma and used without further purification. In the concentration range explored, SDS self-assemble into micelles with an aggregation number of 50 [11]. The nanoparticles investigated were a dispersion of cerium oxide nanocrystals (nanoceria). The nanoceria consisted of isotropic agglomerates of 2 - 5 crystallites with typical size 2 nm and faceted morphologies [12,13]. Their average radius determined by cryo-TEM amounts to 7 nm, with a polydispersity of 0.15. The polydispersity was defined as the ratio between the standard deviation of the size distribution and the average diameter. The particles were coated by poly(acrylic acid) oligomers with molecular weight 2000 g mol$^{-1}$, in a process described recently [12]. The hydrodynamic sizes found in CeO$_2$-PAA$_{2K}$ dispersions was D$_H$ = 13 nm, i.e. 3 nm above that of the bare particles [13]. This 3 nm-increase was assigned to the presence of a strongly charged PAA$_{2K}$ brush around the particles. Electrostatic coacervate complexes were prepared by simple mixing of stock nanoparticle (or surfactant) and copolymer solutions [14,15]. The relative amount of each component was monitored by the charge ratio, which was fixed to unity. Doing so, the solutions were characterized by the same number densities of positive and negative chargeable ions [9,16]. For SDS micelles, the structural charge was assimilated to the aggregation number, yielding a charge of -50e, where e denotes the elementary charge. For the PAA$_{2K}$-coated cerium particles, we have estimated the structural charge of -700e, assuming that 50 % of the 50 PAA$_{2K}$ (per particle) were actually adsorbed at the cerium oxide interface [12].

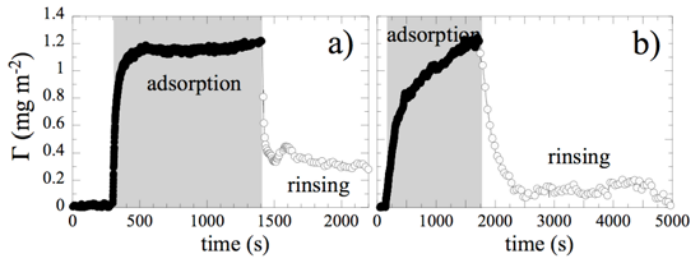

**Figure 2** : *Adsorption kinetics for SDS/PTEA$_{11K}$-b-PAM$_{30K}$ on anionically charged silica (a) and on poly(styrene) (b) substrates, as received from stagnation point adsorption reflectometry (SPAR). Measurements were performed with dilute solutions at c = 0.1 wt. %. Upon rinsing with deionized water, the reflectometry signal dropped down to a level comprised between 0.2 – 0.4 mg m$^{-2}$, indicating an almost complete removal of the layer.*

Static and dynamic light scattering was used to investigate the microstructure and stability of the electrostatic complexes in bulk solutions. The experimental device and data treatment were described recently in Refs. [17] and we refer to this paper for a detailed description. Fig. 1 displays the concentration dependences of the scattering intensities, expressed in terms of the Rayleigh ratio for SDS/PTEA$_{11K}$-b-PAM$_{30K}$ and CeO$_2$-PAA$_{2K}$/PTEA$_{11K}$-b-PAM$_{30K}$. The electrostatic complexes were diluted by addition of de-ionized water at neutral pH. With decreasing concentration, the Rayleigh ratios exhibited a linear dependence in agreement with the dilution law of colloidal systems [18]. For SDS/PTEA$_{11K}$-b-PAM$_{30K}$, below c = 4×10$^{-2}$ wt. %, the scattering intensity dropped abruptly by 2 to 3 orders of magnitudes, attaining then a second linear dependence. By analogy with the micellization of surfactants [19], this second change of slope can be interpreted in terms of a critical association concentration [20,21] (cac), which is here c$_{cac}$ = 6.8×10$^{-3}$ wt. %. Also shown in Fig. 1 are the scattering intensity (thick lines) and the critical micellar concentration of SDS (arrow) calculated if surfactants and polymers were not co-assembled. The good agreement between the intensities in the low concentration range and the calculated Rayleigh ratio suggests that the organic complexes have disassembled below the cac. For the hybrid system CeO$_2$-PAA$_{2K}$/PTEA$_{11K}$-b-PAM$_{30K}$ on the contrary, the scattering remains linear over the whole concentration range. These findings suggest that the cac for the cerium-based aggregates should be lower than 10$^{-4}$ wt. %. Dynamic light scattering has revealed a slightly polydisperse diffusive relaxation mode associated with hydrodynamic diameters D$_H$ = 60 nm for SDS/PTEA$_{11K}$-b-PAM$_{30K}$ (down to c = 2×10$^{-2}$ wt. %) and D$_H$ = 100 nm CeO$_2$-PAA$_{2K}$/PTEA$_{11K}$-b-PAM$_{30K}$. Note that the 100 nm-hybrid aggregates could be identified down to the lowest concentrations.

The adsorption properties of the organic and hybrid complexes on silica and polystyrene substrates were investigated by stagnation point adsorption reflectometry, or SPAR. Experiments were carried out above the cac (c = 0.1 wt. %). A detailed description has been given in the recent literature in Refs. [7,22]. Figs. 2a and 2b show the adsorbed amounts as a function of time for SDS/PTEA$_{11K}$-b-PAM$_{30K}$ system on anionically charged silica and polystyrene substrates, respectively. For both surfaces, adsorption occurred at a level (1.2 mg m$^{-2}$ at steady state) which was consistent with results obtained on electrostatic coacervate phases [23-25]. Upon rinsing (second period of the time traces), the reflectometry signal was found to drop significantly in both cases, although for silica the final amount lied higher that for poly(styrene). These results indicate that the pre-adsorbed layer has been partially removed in the rinsing phase. With the SPAR technique, the addition of deionized water at the stagnation point acts primarily as a dilution process of the deposited layer. As seen for the bulk properties, for the present system, dilution down to below the cac leads to a disassembling of the complexes, and therefore to its almost complete desorption. During the rinsing phase, several scenarios leading to desorption can occur [26]. For silica surfaces, it is plausible that the cationic PTEA blocks adsorb onto the oppositely charged substrate, leading to a finite adsorbed amount, even at long times. In the case of poly(styrene), the removal of the adsorbed materials is almost complete.

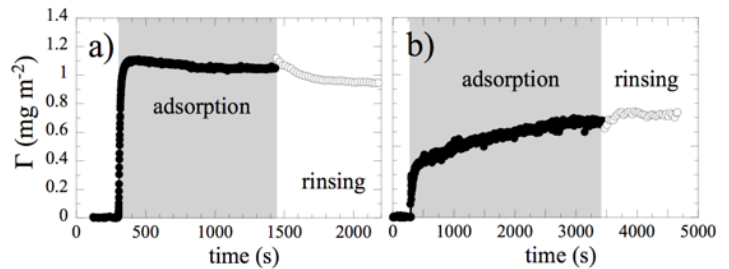

**Figure 3** : *Adsorption kinetics for CeO$_2$-PAA$_{2K}$/PTEA$_{11K}$-b-PAM$_{30K}$ polymer-nanoparticle hybrids on silica (a) and on poly(styrene) (b) substrates. In contradistinction with the data in Fig. 2, the deposited layer did not desorb upon rinsing.*

Figs. 3a and 3b illustrate the adsorption properties of the hybrid coacervates CeO$_2$-PAA$_{2K}$/PTEA$_{11K}$-b-PAM$_{30K}$ on silica and on poly(styrene) substrates, respectively. The two sets of data exhibit a similar behavior. After an adsorption process at a level comparable to those found for the organic complexes (0.8 – 1.5 mg m$^{-2}$), no desorption occurred upon rinsing [27]. We explain the absence of desorption as the result of the stability of the polymer-particle hybrids upon dilution, as illustrated in Fig. 1. Clearly, the existence of a very low cac for the particle-polymer hybrids ensured the stability of the layer deposited on the two substrates. Although weaker, the adsorbed amount of CeO$_2$-PAA$_{2K}$/PTEA$_{11K}$-b-PAM$_{30K}$ on

poly(styrene) was not negligible. After dipping the substrate into a 0.1 wt. % hybrid solution it became hydrophilic, as evidenced by the formation of a wetting film after the removal from the solution and further by a measure of the contact angle lower than 30° (receding conditions).

In conclusion, we have shown how solution properties (here the existence of a very low critical aggregation concentration of nanoparticle/polymer hybrids) can be turned into a valuable surface property, namely the reliability of a deposited layer at a liquid-solid interface. For the present system, we suggest that electrostatic interactions between nanoparticles and oppositely charged polyelectrolytes have resulted in the formation of cores with metastable trapped structures. Because these structures are frozen, their disassembling was prevented even at low concentration. Beyond the sole durability of the coating offered by the hybrid structures, the presence of inorganic nanoparticles carrying their intrinsic properties (anti-UV, high dielectric constant, catalytic effect…) should permit the elaboration of coatings with multiple all-in-one functionalities.

TOC Figure

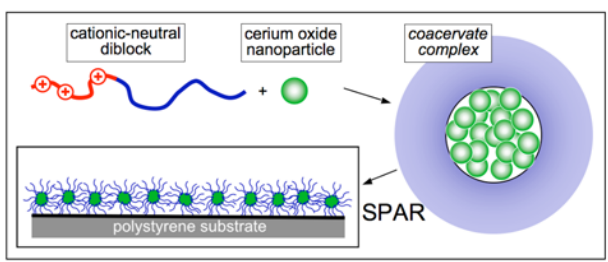